\documentclass[letterpaper, 10 pt, conference]{ieeeconf}  

\IEEEoverridecommandlockouts                              

\usepackage{wrapfig}

\usepackage[english]{babel}
\usepackage{graphicx}
\usepackage{epstopdf}

\usepackage{amsthm}
\usepackage{amsfonts}
\usepackage{amsmath}

\newtheorem{proposition}{Proposition}
\newtheorem{definition}{Definition}
\newtheorem{problem}{Problem}

\newtheorem{theorem}{Theorem}

\usepackage[letterpaper,left=0.75in,right=0.75in,top=0.75in,bottom=1in, twoside=false,includehead,includefoot]{geometry}
\usepackage[english]{babel}

\usepackage{setspace}
\usepackage{tikz}
\usepackage{algorithm}
\usepackage{algpseudocode}
\usepackage{caption}
\usepackage{color}
\usepackage{mathrsfs}

\usepackage{array}
\usepackage{tabularx}
\usepackage[flushleft]{threeparttable}
\usepackage{url}

\newcommand{\upperRomanNum}[1]{\uppercase\expandafter{\rumannumeral#1}}
\newcommand{\lowerRomanNum}[1]{\romannumeral#1\relax}

\usetikzlibrary{automata, positioning, shapes}

\title{\LARGE \bf
Correct-by-synthesis reinforcement learning with temporal logic constraints
}

\author{Min Wen, R\"udiger Ehlers and Ufuk Topcu
\thanks{The work is supported partly by the AFRL,  AFOSR grant \#
FA9440-12-1-0302, ONR grant \# N000141310778 and NSF grant \# 1446479. M. Wen and U. Topcu are with the Department of Electrical and Systems Engineering, University of Pennsylvania, Philadelphia, Pennsylvania, United States. R. Ehlers is with University of Bremen and DFKI GmbH, Bremen, Germany}
}

\begin{document}

\maketitle
\thispagestyle{empty}
\pagestyle{empty}

\begin{abstract}
We consider a problem on the synthesis of reactive controllers that optimize some a priori unknown performance criterion while interacting with an uncontrolled environment such that the system satisfies a given temporal logic specification. We decouple the problem into two subproblems. First, we extract a (maximally) permissive strategy for the system, which encodes multiple (possibly all) ways in which the system can react to the adversarial environment and satisfy the specifications. Then, we quantify the a priori unknown performance criterion as a (still unknown) reward function and compute an optimal strategy for the system within the operating envelope allowed by the permissive strategy by using the so-called maximin-Q learning algorithm. We establish both correctness (with respect to the temporal logic specifications) and optimality (with respect to the a priori unknown performance criterion) of this two-step technique for a fragment of temporal logic specifications. For specifications beyond this fragment, correctness can still be preserved, but the learned strategy may be sub-optimal. We present an algorithm to the overall problem, and demonstrate its use and computational requirements on a set of robot motion planning examples. 
\end{abstract}

\section{INTRODUCTION}

The goal of this paper is to synthesize optimal reactive strategies for systems with respect to some \textit{unknown} performance criterion and in an adversarial environment such that given temporal logic specifications are satisfied. The consideration of unknown performance criterion may seem unreasonable at first sight, but it turns out to be an effective supplement to the specification as task description and suits the need in many applications. On the one hand, general requirements on system behaviors such as safety concerns and task rules may be known and expressed as specifications in temporal logic. On the other hand, quantitative performance criterion can help encode more subtle considerations, such as specific intentions for the current application scenario and personal preferences of human operators who work with the autonomous system. For a path planner of autonomous vehicles, specifications imply fixed nonnegotiable constraints like safety requirements, e.g., always drive on the correct lane, never jump the red light and eventually reach the destination. Quantitative performance criteria give preferences within the context constrained by the specifications, which may involve considerations that have not been taken into account during controller design and suggested by the human operators.

The two main topics most relevant to our work are reactive synthesis with temporal logic specifications and reinforcement learning with respect to unknown performance criteria. Neither solves the problem we consider in this paper. 

On the synthesis side, early work focused on planning in static known environments \cite{manna1984synthesis,vardi1996automata}. Reactivity to the changes in dynamic environments is a crucial functionality. For example, the environment of an autonomous vehicle involves the other vehicles and pedestrians moving nearby, and it is impractical to expect an autonomous vehicle to run on roads safely without reacting to its surrounding environment in real time. Recently, references \cite{piterman2006synthesis,pnueli1989synthesis2,pnueli1989synthesis} considered possibly adversarial environments and reactive strategies (without any quantitative performance criteria). 

Another concern in synthesis is optimality with respect to a given performance criterion. Optimal strategies have been studied with respect to given objectives while satisfying some temporal logic specifications, mostly in deterministic environments or stochastic environments with known transition distribution \cite{ding2014optimal,wolff2012optimal}. Both qualitative objectives such as correctness guarantee with respect to an adversarial environment and quantitative objectives such as mean payoffs were studied in \cite{chatterjee2005mean} though these results crucially rely on the quantitative measure being known a priori.

In order to deal with problems with a priori unknown performance criterion, it is intuitive to gain experience from direct interactions with the environment or with a human operator, which coincides with the motivation of many reinforcement learning methods \cite{sutton1998introduction}. Multiple learning methods have been studied and are available to problems with unknown reward functions and incomplete prior knowledge on system models \cite{barto1983neuronlike,schwartz1993reinforcement,sutton1988learning,watkins1992q}, and have been used in many applications, including the famous TD-Gammon example \cite{tesauro1994td} and robot collision avoidance \cite{huang2005reinforcement}. However, the learning process generally cannot guarantee the satisfaction of other independently imposed specifications while maximizing the expected rewards at the same time, though they can be modified to deal with some simple cases \cite{fu2014probably,perkins2001lyapunov}.

To the best of our knowledge, the current paper is the first to deal with the problem of synthesizing a controller which optimizes some a priori unknown performance criterion while interacting with an uncontrolled environment in a way that satisfies the given temporal logic specifications. The approach we take is based on a decomposition of the problem into two subproblems. For the first part (Section~\ref{sec_permissive_strategies}), the intuition is to extract a strategy for the system, namely a permissive strategy \cite{bernet2002permissive}, which encodes multiple (possibly all) ways in which the system can react to the adversarial environment and satisfy the specifications. Then in the second part (Section~\ref{sec_maxmin_q_learning}), we quantify the a priori unknown performance criterion as a (still unknown) reward function and apply the idea of reinforcement learning to choose an optimal strategy for the system within the operating envelope allowed by the permissive strategy. By decoupling the optimization problem with respect to the unknown cost from the synthesis problem, we manage to synthesize a strategy for the system that is guaranteed to both satisfy the specifications and reach optimality over a set of winning strategies with respect to the a priori unknown performance criterion (Section~\ref{sec_connecting_all_parts}).

\section{Preliminaries}
We now introduce some basic concepts used in the rest.

\subsection{Two-player games}
First we model the {setting }\color{black} as a two-player game. In this model we care about not only the controlled system, but also its external uncontrolled environment. Interactions between the controlled system and the uncontrolled environment play a critical role in guaranteeing the correctness of given specifications, as we will discuss later.

\begin{definition}
A \textit{two-player game}, or simply a \textit{game}, is defined as a tuple $\mathcal{G} = (S, S_s, S_e, I, A_c, A_{uc}, T, W)$, where $S$ is a finite set of states; $\{S_s, S_e\}$ is a partition of $S$, i.e., $S = S_s \bigcup S_e$, $S_s \bigcap S_e = \emptyset$; $I \subseteq S$ is a set of initial states; $A_c$ is a finite set of controlled actions of the system; $A_{uc}$ is a finite set of uncontrolled actions for the environment and $A_{uc} \bigcap A_c = \emptyset$; $T:  S \times \{A_c \bigcup A_{uc}\} \rightarrow 2^S$ is a transition function; $W$ is the winning condition defined later.
\end{definition}

$S_s$ and $S_e$ are the sets of states from which it is the system's or the environment's turn to take actions, respectively. There are no available uncontrolled actions (environment actions) to any state $s \in S_s$, and correspondingly, states in $S_e$ can not respond to any controlled action (system action). Let $A(s)$ be the set of actions available at state $s \in S$. Hence $A(s) \subseteq A_c$ if $s \in S_s$ and $A(s) \subseteq A_{uc}$ if $s \in S_e$.

If the transition function $T$ of $\mathcal{G}$ satisfies $|T(s,a)| \leq 1$ for all $s \in S$ and $a \in A(s)$, the game is called \textit{deterministic}; otherwise the game is called \textit{non-deterministic}, highlighting the fact that multiple actions can be available to some states. We assume here that $\mathcal{G}$ is deterministic.

A \textit{run} $\pi = s_0 s_1 s_2 \ldots$ of $\mathcal{G}$ is an infinite sequence of states such that $s_0 \in S$ and for $i \in \mathbb{N}$, there exists $a_i \in A(s_i)$ such that $s_{i+1} = T(s_i, a_i)$. Without loss of generality, assume that all states are reachable from $I$ in $\mathcal{G}$.

\subsection{Linear temporal logic}
We use fragments of linear temporal logic (LTL) to specify the assumptions on environment behaviors and the requirements for the system. LTL can be regarded as a generalization of propositional logic. In addition to logical connectives such as conjunction ($\wedge$), disjunction ($\vee$), negation ($\neg$) and implication ($\rightarrow$), LTL also includes basic temporal operators such as next ($\bigcirc$), until ($\mathcal{U}$), derived temporal operators like always ($\square$) and eventually ($\diamondsuit$), and any (nested) combination of them, like always eventually ($\square \diamondsuit$).

An \textit{atomic proposition} is a Boolean variable (or propositional variable). Suppose $AP$ is a finite set of atomic propositions, then we can construct LTL formulas as follows: (\lowerRomanNum{1}) Any atomic proposition $p \in AP$ is an LTL formula; (\lowerRomanNum{2}) given formulas $\varphi_1$ and $\varphi_2$, $\neg \varphi_1$, $\varphi_1 \wedge \varphi_2$, $\bigcirc \varphi_1$ and $\varphi_1 \mathcal{U} \varphi_2$ are LTL formulas. A formula without any temporal operators is called a \textit{Boolean formula} or \textit{assertion}. A \textit{linear time property} is a set of infinite sequences over $2^{AP}$. 

LTL formulas are evaluated over executions: An \textit{execution} $\sigma = \sigma_0, \sigma_1, \sigma_2, \ldots$ is an infinite sequence of truth assignments to variables in $AP$, where $\sigma_i$ is the set of atomic propositions that are $True$ at position $i \in \mathbb{N}$. Given an execution $\sigma$ and an LTL formula $\varphi$, the conditions that $\varphi$ \textit{holds at position $i$ of $\sigma$}, denoted by $\sigma, i \models \varphi$, are constructed inductively as follows: (\lowerRomanNum{1}) Let $P(\varphi)$ be the set of atomic propositions appearing in $\varphi$. Then for any $p \in P(\varphi)$, $\sigma, i \models p$ iff $p \in \sigma_i$. (\lowerRomanNum{2}) $\sigma, i \models \neg \varphi$ iff $\sigma, i \not \models \varphi$. (\lowerRomanNum{3}) $\sigma, i \models \bigcirc \varphi$ iff $\sigma, i+1 \models \varphi$. (\lowerRomanNum{4}) If $\varphi = \varphi_1 \wedge \varphi_2$, then $\sigma, i \models \varphi$ iff $\sigma, i \models \varphi_1$ and $\sigma, i \models \varphi_2$. (\lowerRomanNum{5}) If $\varphi = \varphi_1 \vee \varphi_2$, then $\sigma, i \models \varphi$ iff $\sigma, i \models \varphi_1$ or $\sigma, i \models \varphi_2$. (\lowerRomanNum{6}) If $\varphi = (\varphi_1 \rightarrow \varphi_2)$, then $\sigma, i \models \varphi$ iff $\sigma, i \models \varphi_1$ implies $\sigma, i \models \varphi_2$.(\lowerRomanNum{7}) If $\varphi = \varphi_1 \mathcal{U} \varphi_2$, then $\sigma, i \models \varphi$ iff there exists $k \geq i$ such that $\sigma, j \models \varphi_1$ for all $i \leq j < k$ and $\sigma, k \models \varphi_2$. (\lowerRomanNum{7}) $\diamondsuit \varphi = $ \textit{True} $\mathcal{U} \varphi$, $\square \varphi = \neg \diamondsuit \neg \varphi$.
We say that $\varphi$ holds on $\sigma$ or $\sigma$ satisfies $\varphi$, denoted by $\sigma \models \varphi$, if $\sigma, 0 \models \varphi$.

An LTL formula $\varphi_1$ is a \textit{safety formula} if for every execution $\sigma$ that violates $\varphi_1$, there exists an $i \in \mathbb{N}^+$ such that for every execution $\sigma^{\prime}$ that coincides with $\sigma$ up to position $i$, $\sigma^{\prime}$ also violates $\varphi_1$. An LTL formula $\varphi_2$ is a \textit{liveness formula} if for every prefix of any execution  $\sigma_0, \ldots, \sigma_i$ $(i \geq 0)$, there exists an infinite execution $\sigma^{\prime}$ with prefix $\sigma_0, \ldots, \sigma_i$ such that $\sigma^{\prime} \models \varphi_2$. Intuitively, safety formulas indicate that ``something bad should never happen,'' and liveness formulas require that ``good things will happen eventually.''

Let $AP$ be a set of atomic propositions, and define a \textit{labeling function} $L: S \rightarrow 2^{AP}$ such that each state $s \in S$ is mapped to the set of atomic propositions that hold $True$ at state $s$. A \textit{word} is an infinite sequence of labels $L(\pi) = L(s_0) L(s_1) L(s_2) \ldots$ where $\pi = s_0 s_1 s_2 \ldots$ is a run of $\mathcal{G}$. We say a run $\pi$ satisfies $\varphi$ if and only if $L(\pi) \models \varphi$.

To complete the definition of two-player games, define the winning condition $W = (L, \varphi)$ such that $L$ is a labeling function and $\varphi$ is an LTL formula, and a run $\pi$ of $\mathcal{G}$ is \textit{winning for the system} if and only if $\pi$ satisfies $\varphi$. $\varphi$ can be used to express the qualitative specifications such as system requirements and environment assumptions.  \vspace{-0.04in}

\subsection{Control strategies}
Given the game $\mathcal{G}$, we would like to synthesize a control protocol such that the runs of $\mathcal{G}$ satisfy the specification $\varphi$.

A \textit{(deterministic) memoryless strategy for the system} is a map $\mu: S_s \rightarrow A_c$, where $\mu(s) \in A(s)$ for all $s \in S_s$. A \textit{(deterministic) finite-memory strategy for the system} is a tuple $\mu = (\mu_m, \rho_m, M)$ where $\mu_m: S_s \times M \rightarrow A_c$ such that $\mu_m(s,m) \in A(s)$ for all $s \in S_s$, $m \in M$, and $\rho_m: S \times M \rightarrow M$. The finite set $M$ is called the \textit{memory} and $\rho_m$ is also called the \textit{memory update function}. $\mu_m(s, m) \in A(s)$ for all $s \in S_s$ and $m \in M$. $m$ is initialized to be $m_0 \in M$. Strategies can also be defined as \textit{non-deterministic}, \color{black} in which case \color{black} $\mu$ will be defined as $\mu: S_s \rightarrow 2^{A_c}$ for memoryless strategies or $\mu = (\mu_m, \rho_m, M \color{black})$ with $\mu_m: S_s \times M \rightarrow 2^{A_c}$ for finite-memory strategies. Clearly deterministic strategies can be regarded as a special case of non-deterministic strategies when $M$ is a singleton. We require $|\rho_m(s, m)| = 1$ for all $s \in S$ and $m \in M$, no matter the strategy is deterministic or not. $\rho_m$ will be evaluated each time after any player takes action. If we further specify the probability distribution ${P}$ over $A(s)$ for each state $s \in S_s$, the corresponding strategies are called \textit{randomized strategies}. We refer to deterministic strategies unless otherwise stated. By replacing $S_s$ by $S_e$ and $A_c$ by $A_{uc}$, we can define memoryless and finite-memory strategy for the environment.

A run $\pi = s_0 s_1 s_2 \ldots$ is \textit{induced by a strategy} $\mu$ \textit{for the system} if for any $i\in \mathbb{N}$ such that $s_i \in S_s$, $s_{i+1}=T(s_i, \mu(s_i))$ (for memoryless strategies) or there exists an infinite sequence of memories $m_0 m_1 m_2 \ldots \in M^{\omega}$ such that $s_{i+1}=T(s_i,\mu_m(s_i, m_i))$ and for all $s_j \in S$, $m_{j+1} = \rho_m(s_{j+1}, m_j)$ (for finite-memory strategies). 
Let ${R}^{\mu}(s)$ be the set of runs of $\mathcal{G}$ induced by a strategy $\mu$ for the system and initialized with $s \in S$. $|R^{\mu}(s)| > 1$ when the strategies for the environment are not unique, even if $\mu$ is deterministic.

We say a strategy $\mu$ for the system $\textit{wins at state }s \in S$ if all runs $\pi \in {R}^{\mu}(s)$ are winning for the system. A strategy $\mu$ is called a \textit{winning strategy} if it wins at all initial states of $\mathcal{G}$. A formula $\varphi$ is \textit{realizable} for $\mathcal{G}$ if there exists a winning strategy for the system with $W = (L, \varphi)$. 

\subsection{Reward functions}
Besides qualitative requirements which are encoded in the winning condition, we also consider quantitative evaluation from other sources such as the human operators. Such evaluation is modeled as a reward function which we want to maximize by choosing proper strategy for the system.

In order to evaluate the system strategy, we first map each system state-action pair to a nonnegative value by an \textit{instantaneous reward function} $\mathcal{R}: S_s \times (A_c \bigcup A_{uc}) \rightarrow \mathbb{R}^+ \bigcup \{0\}$, and then consider the ``accumulation'' of such instantaneous rewards obtained over a run of a game $\mathcal{G}$. As runs are of infinite length, we cannot simply add all the instantaneous reward acquired, which may approach infinity. Instead we define a \textit{reward function} $J^{\mathcal{G}}_{\mathcal{R}}: S^{\omega} \rightarrow \mathbb{R}$ to compute reward for any run $\pi$ of $\mathcal{G}$ given the instantaneous reward function $\mathcal{R}$. A common example of $J^{\mathcal{G}}_{\mathcal{R}}$ is the discounted reward
\begin{equation}
\label{eq:discounted}
J^{\mathcal{G}}_{\mathcal{R}} = \sum_{k=0}^{\infty} \gamma^k R_{{k+1} \color{black}},
\end{equation}
where $\gamma$ is a discount factor satisfying $0 \leq \gamma < 1$, and $R_{k+1}$ is the $(k+1)$th instantaneous reward obtained by the system. In this case, rewards acquired earlier are given more weight, while in other examples like the mean payoff function $\liminf_{k\rightarrow \infty}\frac{1}{k+1} \sum_{i=t}^{t+k} R_{i}$, weights on instantaneous reward are independent of the sequence.

Now we define a reward function $\bar{J}^{\mathcal{G}}_{\mathcal{R}}: \mathcal{P} \times S_s \rightarrow \mathbb{R}^+ \bigcup \{0\}$ to evaluate each strategy for the system, where $\mathcal{P}$ is the set of system strategies. Usually $|R^{\mu}(s)| > 1$ as the uncontrolled environment has more than one strategies, and thus the definition of $\bar{J}^{\mathcal{G}}_{\mathcal{R}}(\mu, s)$ is not unique given $J^{\mathcal{G}}_{\mathcal{R}}(\pi)$ for all runs in $R^{\mu}(s)$. One commonly used choice for $\bar{J}^{\mathcal{G}}_{\mathcal{R}}$ is the expectation of $J^{\mathcal{G}}_{\mathcal{R}}(\pi)$ over all runs in $R^{\mu}(s)$ with some given distribution for the environment strategy, i.e., $\mathbb{E}_{\pi \in R^{\mu}(s)}\big[ J^{\mathcal{G}}_{\mathcal{R}}(\pi) \big].$ The distribution is usually estimated from interaction experience with the environment. Another common way is to define $\bar{J}^{\mathcal{G}}_{\mathcal{R}}$ as the minimal possible reward acquired when the system strategy is $\mu$, i.e., 
\begin{equation}
\bar{J}^{{\mathcal{G}}}_{{\mathcal{R}}}({\mu}, {s}) = \inf_{\pi \in R^{{\mu}}({s})} J^{{\mathcal{G}}}_{{\mathcal{R}}}(\pi),
\label{eq:worst_reward}
\end{equation}
which we use as the reward function in our problem. 

\section{Problem formulation and overview of the solution approach}

We have modeled the interaction between the uncontrolled environment and the controlled system as a two-player game whose winning condition is described by a given LTL formula. Moreover, we defined reward functions to evaluate the performance of different system strategies. Now we can go on to formulate the main problem of the paper.

\begin{problem}
A two-player deterministic game $\mathcal{G} = (S, S_s, S_e, I, A_c, A_{uc}, T, W)$ is given where $W = (L, \varphi)$ and $\varphi$ is realizable for $\mathcal{G}$. Find a memoryless or finite-memory winning strategy $\mu$ for the system such that $\bar{J}^{\mathcal{G}}_{\mathcal{R}}({\mu},s)$ is maximized for all $s \in I$, where a reward function $\bar{J}^{\mathcal{G}}_{\mathcal{R}}$ is given with respect to an unknown instantaneous reward function $\mathcal{R}: S_s \times (A_c \bigcup A_{uc}) \rightarrow \mathbb{R}^+ \bigcup \{0\}$.
\label{problem_overall}
\end{problem}

Generally, there does not necessarily exist a memoryless or finite-memory winning strategy that maximizes the reward over all winning strategies, as it is possible that the instantaneous reward promotes the system to violate the specification. Take as an example $\mathcal{G}_0 = (S, S_s, S_e, I, A_c, A_{uc}, T, W)$, where $S_e = \emptyset$, $S = S_s = I = \{s_0, s_1\}, A_c = \{a_0, a_1, a_2\}, A_{uc} = \emptyset$ and $W = (L, \varphi)$. 
The transition function $T$ and the labeling function $L$ are shown in Fig.~\ref{fig:no_finite}, and the formula is $\varphi = \diamondsuit b_2$.
\begin{wrapfigure}{r}{0.28\textwidth} \vspace{-0.26in}
\begin{center}
\begin{tikzpicture}[shorten >=1pt,node distance=2cm,on grid,auto]
\tikzstyle{every node} = [circle, fill=white]
   \node[state] (s_0)   {$s_0$};
   \node[above=12pt] {$\{b_1\}$} (0,0);
   \node[state] (s_1) at +(1.7,0)  {$s_1$};
   \node[above=12pt] at +(1.7,0) {$\{b_2\}$};
   \path[->]
    (s_0) edge [right] node[above] {$a_1$} (s_1)
    (s_0) edge [loop left] node {$a_0$} (s_0)
    (s_1) edge [loop right] node {$a_2$} (s_1);
\end{tikzpicture}
\caption{A game $\mathcal{G}_0$ without finite-memory optimal strategy.}
\label{fig:no_finite}  \vspace{-0.18in}
\end{center}
\end{wrapfigure}
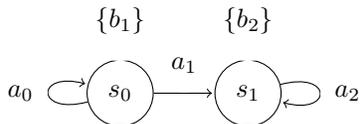  The game $\mathcal{G}_0$ does have winning strategies for the system. For example, the strategy $\mu$ where $\mu(s_0) = a_1, \mu(s_1) = a_2$ is a memoryless winning strategy, and the strategy $\mu^{\prime} = (\mu_m^{\prime},\rho_m^{\prime}, \{0, 1\})$ where $\mu_m^{\prime}(s_1, 0) = \mu_m^{\prime}(s_1, 1) = \{a_2\}$, $\mu_m^{\prime}(s_0, 0) = \{a_0, a_1\}$, $\mu_m^{\prime}(s_0, 1) = \{a_1\}$ and $\rho_m^{\prime}(s_1, 0) = 0$, $\rho_m^{\prime}(s_1, 1) = \rho_m^{\prime}(s_0, 0) = \rho_m^{\prime}(s_0, 1) = 1$, is a finite-memory winning strategy.

But $\mathcal{G}_0$ does not have memoryless or finite-memory optimal winning strategies for the system. Suppose the unknown instantaneous reward function is actually defined as $\mathcal{R}(s_1, a_2) = 0$, $\mathcal{R}(s_0, a_0) = \mathcal{R}(s_0, a_1) = 10$, and the reward function $J^{\mathcal{G}}_{\mathcal{R}}$ is the same as \eqref{eq:discounted}. In order to maximize $\bar{J}^{\mathcal{G}}_\mathcal{R}({\mu},\cdot)$, $\mu$ should allow the system to stay at $s_0$ as many times as possible, but staying at $s_0$ forever will violate $\varphi$. Thus optimal winning strategies need infinite memory.


Let us now move on to an overview of the two-stage solution approach we propose. Given a game $\mathcal{G}$ as in Problem~\ref{problem_overall}, we first extract a non-deterministic winning strategy $\mu_p$ called a \textit{permissive strategy} \cite{bernet2002permissive}, which guarantees that $R^{\mu}(s) \subseteq R^{\mu_p}(s)$ for all memoryless winning strategies $\mu$ and $s \in I$. In some special cases (e.g. the conditions in Proposition~\ref{prop_special_safety}), we are even able to compute a \textit{maximally permissive strategy} $\mu_p^{max}$, such that $R^{\mu}(s) \subseteq R^{\mu_p^{max}}(s)$ for all winning strategies $\mu$ and $s \in I$. Then in the second stage we restrict to the transitions allowed by $\mu_p$ (or $\mu_p^{max}$), apply reinforcement learning methods to explore the a priori unknown instantaneous reward function $\mathcal{R}$ and compute an optimal strategy over all strategies of the new game obtained in the first stage. With this decomposition we managed to separate the problem of guaranteeing the correctness of specifications from that of seeking the optimality of the reward function with a priori unknown instantaneous rewards. \vspace{-0.1in}


\section{Permissive strategies, learning and the main algorithm}
This section is composed of three parts. We first introduce the idea of permissive strategies, then describe a reinforcement learning method which is used to learn an optimal strategy with respect to an unknown reward function without concern about any specification, and finally combine the two parts to apply the reinforcement learning method to explore for an optimal strategy out of those encoded in an appropriately constructed permissive strategy.

\subsection{Extraction of permissive strategies}
\label{sec_permissive_strategies}

 
We first introduce an inclusion relation between strategies. Recall that we have defined the set of runs induced by a strategy $\mu$ for the system with initial state $s \in I$ as $R^{\mu}(s)$. For two non-deterministic strategies $\mu_1$ and $\mu_2$ for the system, we say that $\mu_1$ \textit{includes} $\mu_2$ if $R^{\mu_2}(s) \subseteq R^{\mu_1}(s)$ holds for all $s \in I$. Furthermore, if $\mu_1$ includes $\mu_2$ and $\mu_2$ includes $\mu_1$, we call $\mu_1$ and $\mu_2$ \textit{equivalent}. In other words, equivalent strategies induce the same set of runs. A game $\mathcal{G}$ has a \textit{unique} winning strategy if all its winning strategies are equivalent. Now we can define permissive strategies based on this strategy inclusion relation.

\begin{definition}
Given a two-player game $\mathcal{G}$, a non-deterministic strategy $\mu$ for the system is called \textit{permissive} if (\lowerRomanNum{1}) it is winning for the system and (\lowerRomanNum{2}) includes all memoryless winning strategies for the system. A permissive strategy is called \textit{maximally permissive} if it includes all winning strategies for the system.
\label{def_permissive}
\end{definition}

All two-player games have permissive strategies. For games with finite states, there are only finitely many memoryless winning strategies. We can build a permissive strategy by adding a unique tag to each of them (as memory) and directly combining them together. In cases where there is no memoryless winning strategy, this fact is trivial as any winning strategy is permissive. 

In general, permissive strategies are not necessarily unique. 
For example, the game $\mathcal{G}_0$ in Fig.~\ref{fig:no_finite} has a unique memoryless winning strategy $\mu$ for the system where $\mu(s_0) = a_1$, $\mu(s_1) = a_2$. As a result, $\mu_p$ such that $\mu_p(s_0) = \{a_1\}$ and $\mu_p(s_1) = \{a_2\}$ is a deterministic memoryless permissive strategy. 
In the meantime, the finite-memory strategy  $\mu^{\prime} = (\mu_m^{\prime},\rho_m^{\prime}, \{0,1\} \color{black})$ where $\mu_m^{\prime}(s_1, 0) = \mu_m^{\prime}(s_1, 1) = \{a_2\}$, $\mu_m^{\prime}(s_0, 0) = \{a_0, a_1\}$, $\mu_m^{\prime}(s_0, 1) = \{a_1\}$ and $\rho_m^{\prime}(s_0, 0) = \rho_m^{\prime}(s_0, 1) = \rho_m^{\prime}(s_1, 1) = 0$, $\rho_m^{\prime}(s_1, 0) = 0$ includes $\mu$ and thus is also a permissive strategy. As $\mu_p$ does not include $\mu^{\prime}$, they are different permissive strategies of $\mathcal{G}$.

On the other hand, maximally permissive strategies must be unique by definition, if they exist for a game $\mathcal{G}$. The specific representations of maximally permissive strategies may not be unique, just like a memoryless strategy can be rewritten as a finite-memory strategy in which the allowed actions are independent of the memory.

It is naturally desirable to extract maximally permissive strategies as they include all the other winning strategies. While they do not exist in general, the following proposition is a sufficient condition of their existence.
\begin{proposition}[\cite{bernet2002permissive}]
All games $\mathcal{G}$ with winning conditions $W=(L,\varphi)$ in which $\varphi$ is a safety formula have memoryless or finite-memory maximally permissive strategies. 
\label{prop_safety_permissive}
\end{proposition}
This characterization can be extended to be both sufficient and necessary. It has been shown that precisely for the \emph{reactive safety properties} \cite{DBLP:journals/corr/abs-1106-1240}, maximally permissive strategies exist. These are the properties equivalent to safety properties when the interaction between the environment and system is explicitly considered, i.e., reactive safety characterizes precisely the properties whose satisfaction can be checked by testing if the runs of $\mathcal{G}$ satisfy some safety formula.

There exists work on the construction of permissive strategies for games $\mathcal{G}$ with a general LTL formula $\varphi$ in the winning condition  \cite{bernet2002permissive,DBLP:journals/sttt/SohailS13}, so we only sketch the relevant results briefly here. 
The first step is to compute a \emph{deterministic parity automaton} \cite{thomas2002automata} from $\varphi$, which is taken into account by constructing a new game $\mathcal{G}'$. $\mathcal{G}'$ uses a \emph{parity winning condition} for $\varphi'$, which is of the form $\bigwedge_{0 \leq c \leq n} (\square \diamondsuit K_{2 \cdot i} \wedge \diamondsuit \square K_{2 \cdot i+1})$ for some state formulas $K_1, \cdots, K_{2n+1}$ that mutually imply each other, i.e., for which for all $0 \leq i \leq 2n$, $K_{i+1} \rightarrow K_i$ is a valid LTL formula \cite{DBLP:journals/sttt/SohailS13}. Games with such a winning condition  have permissive strategies and Bernet et al. \cite{bernet2002permissive} provided a construction method to compute such strategies.
Additionally, Ehlers and Finkbeiner \cite{DBLP:journals/corr/abs-1106-1240} offer a method for checking if the winning condition $W$ is a reactive safety property for $\mathcal{G}$. 
The game $\mathcal{G}$ is first translated into a parity automaton $\mathscr{A}$ as before, and is then used to construct a parity tree automaton $\mathscr{A}'$ \cite{thomas2002automata}. Tree automata are commonly used to explicitly model inputs and outputs and the overall behavior of reactive systems. 
For reactive safety properties, trees get rejected if and only if some path in the tree visits some violating states, i.e., the states from which all trees are rejected. In other words, the set of accepted trees should be exactly those that never visit any violating state in all paths. The acceptance of trees can be decided by simply checking the set of states they can visit. The problem of checking if $\varphi$ is a reactive safety property for $\mathscr{A}$ is reduced to checking the equivalence of two parity tree automata, which can be solved with existing approaches \cite{jurdzinski2000small}. If we get a positive answer, we can construct another game with a safety formula in its winning condition which accepts exactly the same set of runs as $\mathscr{A}$. By Proposition~\ref{prop_safety_permissive}, there exists a maximally permissive strategy for $\mathcal{G}$. The worst-case complexity of the resulting method is 2-EXPTIME.  

Although Proposition~\ref{prop_safety_permissive} guarantees the existence of a maximally permissive strategy $\mu_p^{max}$ when $\varphi$ is a safety formula,  the computational time complexity is the same as that of synthesizing a strategy for a game with a general LTL formula, which is 2-EXPTIME \cite{kupferman2001model}. 
It can be significantly improved when $\varphi$ is of the following special form. The proof is straightforward and is omitted due to the limited space.

\begin{proposition}
For all games $\mathcal{G}$ with winning condition $W = (L, \varphi)$ and $\varphi = \varphi_0 \wedge \Box \varphi_1$, where $\varphi_0$ and $\varphi_1$ are Boolean formulas of $p$ and $\bigcirc q$ for $p,q \in AP$, a memoryless maximally permissive strategy can be solved in linear time of the number of transitions of $\mathcal{G}$ and the size of $\varphi$.
\label{prop_special_safety}
\end{proposition}
 



We use the software tool \texttt{slugs} \cite{web_slugs} to extract permissive strategies when $\varphi$ in $\mathcal{G}$ is in the form of generalized reactivity (1) (GR(1)) \cite{piterman2006synthesis}. 
Under the condition of Proposition~\ref{prop_special_safety}, \texttt{slugs} synthesizes a maximally permissive strategy.

The extraction and application of permissive strategies greatly simplify the solution of Problem~\ref{problem_overall}, enabling us to focus on optimizing the  performance through strategies known to be correct. By Definition~\ref{def_permissive}, a permissive strategy $\mu_p$ is non-deterministic and thus its application to a game $\mathcal{G}$ is essentially encoding its memory update function into the game structure and removing all transitions that it does not allow. Hence any run $\pi'$ of the resulting game $\mathcal{G}'$ has a unique counterpart $\pi$ in the runs of $\mathcal{G}$ induced by $\mu_p$, and vice versa. Moreover, such $\pi$ and $\pi'$ can only be winning for the system simultaneously. 
Since $\mu_p$ is winning for the system in $\mathcal{G}$, all runs it induces are winning for the system and so are their counterpart runs in $\mathcal{G}'$. As a result, any strategy $\mu'$ of $\mathcal{G}'$ is winning for the system. Let $J^{\mathcal{G}'}_{\mathcal{R}}(\pi')$ be the same as $J^{\mathcal{G}}_{\mathcal{R}}(\pi)$, and $\bar{J}^{\mathcal{G}'}_{\mathcal{R}}$ is defined similarly as $\bar{J}^{\mathcal{G}}_{\mathcal{R}}$.


\subsection{Reinforcement learning}
\label{sec_maxmin_q_learning}
Now that we have acquired a game $\mathcal{G}^{\prime}$ whose runs are all guaranteed to be correct with respect to the underlying linear temporal logic specification, we can move on to learn an optimal strategy with respect to an a priori unknown instantaneous reward function $\mathcal{R}$. The reinforcement learning algorithm aim to maximize $\bar{J}^{\mathcal{G}'}_{\mathcal{R}}$ for the game $\mathcal{G}'$. 

The choice of reinforcement learning algorithms depends on the choice of the reward function $\bar{J}^{\mathcal{G}}_{\mathcal{R}}$ in Problem~\ref{problem_overall}, regardless of how the permissive strategy $\mu_p$ is generated. Here we focus on discounted reward functions, but the pseudo-algorithm in Section~\ref{sec_connecting_all_parts} also works with other forms of $\bar{J}^{\mathcal{G}'}_{\mathcal{R}}$ so long as there exists an optimal deterministic winning strategy $\mu'$ which can be solved by the corresponding reinforcement learning method. 

The discounted reward function for evaluating the rewards obtained by a run is shown in \eqref{eq:discounted}. We particularly focus on the minimal (worst-case) possible reward for each system strategy, as shown in \eqref{eq:worst_reward}. This definition concerns about the tight lower bound of the reward obtained by executing strategy $\mu'$ whatever strategy the uncontrolled environment implements. In other words, we assume that the environment acts adversarially and the game is equivalently a zero-sum game. 
It has been shown that in this case 
both the environment and the system have deterministic memoryless optimal strategies in $\mathcal{G}'$ \cite{shapley1953stochastic}. 
As a result we can neglect all randomized strategies without loss of optimality.
Such an optimal strategy can be computed by the maximin-Q algorithm, which is a simple variation of the minimax-Q learning algorithm \cite{littman1994markov}. It can also be solved by the generalized Q-learning algorithm for alternating Markov games \cite{littman1996generalized}. Both methods guarantee that the learned greedy strategy, which always chooses an action with the best learned Q value, converges to an optimal strategy for a system interacting with an adversary under some common convergence conditions.  \vspace{-0.04in}

\subsection{Connecting the dots: correct-by-synthesis learning}
\label{sec_connecting_all_parts}
Having discussed permissive strategies and reinforcement learning, we are now ready to connect the pieces and discuss a solution to Problem~\ref{problem_overall}, which is outlined in Algorithm~\ref{Algorithm_pseudo}. Maximally permissive strategies play a special role as they include all winning strategies for the system, 
and their existence naturally divide the solution into two cases.

\begin{algorithm}[!ht]
\caption{Pseudo-algorithm for solving Problem~\ref{problem_overall}}
\label{Algorithm_pseudo}
\begin{algorithmic}
\Require {A game $\mathcal{G} = (S, S_e, S_s, I, A_c, A_{uc}, T, W)$ with $W = (L, \varphi)$} in which $\varphi$ is a realizable formula for $\mathcal{G}$, a reward function ${J}^{\mathcal{G}}_{\mathcal{R}}$ and $\bar{J}^{{\mathcal{G}}}_{{\mathcal{R}}}$ (e.g. as in \eqref{eq:discounted}, \eqref{eq:worst_reward}) with respect to an unknown instantaneous reward function $\mathcal{R}$.
\Ensure {A winning strategy $\mu$ for the system that maximizes $\bar{J}^{\mathcal{G}}_{\mathcal{R}}(\mu, s)$ for all $s \in I$.}
\State \textbf{Step 1.} Compute a (maximally) permissive strategy $\mu_p$.
\State \textbf{Step 2.} Apply $\mu_p$ to $\mathcal{G}$ and modify $\mathcal{G}$ into a new game $\hat{\mathcal{G}} = (\hat{S}, \hat{S_s}, \hat{S_e}, \hat{I}, A_c, A_{uc}, \hat{T}, \hat{W})$, where $\hat{W} = (L, True)$. 
\State \textbf{Step 3.} Compute $\hat{\mu}^*$ that maximizes $\hat{J}^{\hat{\mathcal{G}}}_{\mathcal{R}(\mu, s)}$ for all $s \in I$ with some reinforcement learning algorithm (e.g. the maximin-Q algorithm).
\State \textbf{Step 4.} Map $\hat{\mu}^*$ in $\hat{\mathcal{G}}$ back to $\mu^*$ in $\mathcal{G}$.
\State \Return $\mu^*$.
\end{algorithmic}
\end{algorithm}

\vspace{-0.5\baselineskip}
\subsubsection{For games whose maximally permissive strategies can be computed}
If maximally permissive strategies can be computed for a game $\mathcal{G}$, 
$\mu_p$ in Algorithm~\ref{Algorithm_pseudo} includes all winning strategies and is a winning strategy itself. Applying it to $\mathcal{G}$ not only guarantees winning for the system but also preserves all winning strategies for the system in all subsequent steps, which decouples the correctness requirements from optimality concerns. As the output of the reinforcement learning algorithm used in Step 3 is guaranteed to converge to an optimal deterministic winning strategy, the output of Algorithm~\ref{Algorithm_pseudo} is guaranteed to be a solution of Problem~\ref{problem_overall}. 
Theorem~\ref{thm_sufficient_overall} summarizes this result in a special case. 
\begin{theorem}
If the conditions in Proposition~\ref{prop_special_safety} hold, the output of Algorithm~\ref{Algorithm_pseudo} is a solution to Problem~\ref{problem_overall}.
\label{thm_sufficient_overall}
\end{theorem}

\subsubsection{For games whose maximally permissive strategies cannot be computed}
\label{sec_mps_does_not_exist}


If maximally permissive strategies for a game $\mathcal{G}$ are not solvable, the best we can expect is to extract a permissive strategy which includes a proper subset of winning strategies for the system. There can be many permissive strategies for the same game with different ``permissiveness'', i.e.,  including different subsets of winning strategies. For two different permissive strategies $\mu_1$ and $\mu_2$ for the system, if $\mu_2$ includes $\mu_1$, intuitively $\mu_2$ would be more ``permissive'' and have higher worst-case reward, although it is also expected to consume more computation resources. Thus there is a natural trade-off between ``permissiveness'' and optimality for the solution of this case, which is illustrated in Section~\ref{sec_examples}.

\section{Examples}
\label{sec_examples}
We demonstrate the use of Algorithm~\ref{Algorithm_pseudo} on robot motion planning examples in grid worlds with different sizes and winning specifications. The game in the first example has a maximally permissive strategy for the system as its specification is a safety formula, while for the second example we can at most compute a permissive strategy. The last example shows the trade-off between the performance of the learned system strategy of Algorithm~\ref{Algorithm_pseudo} and the computation cost.

{\bf Example 1:} Two robots, namely a system robot and an environment robot, move in an $N$-by-$N$ square grid world strictly in turns. It is known that the two robots are in different cells initially and at each move, the environment robot must go to an adjacent cell, while the system robot can either go to an adjacent cell or stay in its current cell. 
The system robot should always avoid collision with the environment robot. Assume that the positions of both robots are always observable for the system. 

This problem can be formulated as a game $\mathcal{G} = \{S, S_s, S_e, I, A_c, A_{uc}, T, W\}$ with $W = (L, \varphi_0)$. Let $Pos = \{0, \ldots, {N^2-1}\}$ be the set of cells in the map. Then $S = Pos\times Pos \times\{0, 1\}$, $S_s = Pos\times Pos\times\{1\}$, $S_e=Pos\times Pos\times \{0\}$. $I = \{(x,y,1)\mid x,y \in Pos, x \not= y\}$. $A_c = \{up_s, down_s, left_s, right_s, stay_s\}$, $A_{uc} = \{up_e, down_e, left_e, right_e\}$. The transition function $T$ guarantees that $A_c$ and $A_{uc}$ only change the first and second component of a state respectively. The set of atomic propositions is $AP = \big( \bigcup_{i=0}^{{N^2-1}} x_i \big) \cup \big( \bigcup_{j=0}^{{N^2-1}} y_j \big) \cup \{t_0, t_1\}$. The labeling function is $L(s) = \{x_i, y_j, t_k\}$ if $s = (i, j, k) \in S$. The requirements on the system robot can be expressed as $\varphi_0 = \bigwedge\nolimits_{i=0}^{{N^2-1}} (\neg x_i \vee \neg y_i) \wedge \square \bigwedge\nolimits_{i=0}^{{N^2-1}} (x_i \rightarrow \neg y_i).$ 
Proposition~\ref{prop_special_safety} asserts that we can compute a maximally permissive strategy and construct $\hat{\mathcal{G}}$. By Theorem~\ref{thm_sufficient_overall}, Algorithm~\ref{Algorithm_pseudo} is expected to output an optimal strategy for the system.


The reward functions $J^{\mathcal{G}}_{\mathcal{R}}$ and $\bar{J}^{\mathcal{G}}_{\mathcal{R}}$ are given as \eqref{eq:discounted} and \eqref{eq:worst_reward}, with the discounting factor $\gamma$ set to be 0.9. However, the instantaneous reward function $\mathcal{R}$ is a priori unknown to the system robot. In practical scenarios $\mathcal{R}$ is often given by some independent human operator or trainer of the system robot for unpredictable purposes with arbitrarily complicated structure and thus can neither be acquired nor be guessed ahead of time. For this numerical example, $\mathcal{R}$ is set to encourage the system robot to reach positions diagonal to the environment robot's position as often as possible. From a state $s \in S_s$, $\mathcal{R}(s,a) = 1$ if the two robots are diagonal to each other at $T(s,a)$, otherwise $\mathcal{R}(s,a)=0$. But this information is not available to the system robot in advance and is only revealed through the learning process. The system robot can only get an instantaneous reward each time when it takes a corresponding transition.  

\begin{table}[h]
\centering
\caption{Results for example 1.}
\label{table_demo1}
\begin{tabular}{| >{\centering\arraybackslash}m{0.5cm} | >{\centering\arraybackslash}m{1cm} | >{\centering\arraybackslash}m{1cm} | c | c | c | c |} 
\hline
{$N$} & {$t_e$ [s]} & {$t_l$ [s]} & {Iterations} & {$|\hat{S}|$} & {$|\hat{S}_s|$} \\ \hline
3 & 0.10 & 4.28 & $9 \times 10^4$ & 120 & 72 \\ \hline 
4 & 0.21 & 16.35 & $3.2 \times 10^5$ & 432 & 240 \\ \hline
5 & 2.20 & 43.12 & $8.5 \times 10^5$ & 1120 & 600 \\ \hline 
6 & 19.40 & 88.69 & $1.81 \times 10^6$ & 2400 & 1260 \\ \hline 
8 & 30.29 & 305.77 & $6.05 \times 10^6$ & 7840 & 4032 \\ \hline 
10 & 300.00 & 771.73 & $1.562 \times 10^7$ & 19440 & 9900 \\ \hline 
\end{tabular}
\end{table}

The results for the cases when $N = 3, 4, 5, 6, 8, 10$ are shown in Table~\ref{table_demo1}, 
where $t_e$ is the time [s] spent extracting a maximally permissive strategy $\mu_p^{max}$ with \texttt{slugs}, and $t_l$ is the time [s] used to learn an optimal strategy $\hat{\mu}^*$. The number of states and state-action tuples are for the game $\hat{\mathcal{G}}$ in Algorithm~\ref{Algorithm_pseudo}. 
All examples run on a laptop with a 2.4GHz CPU and 8GB memory. 

\begin{figure}[ht!]
\centering
\includegraphics[width=0.4\textwidth]{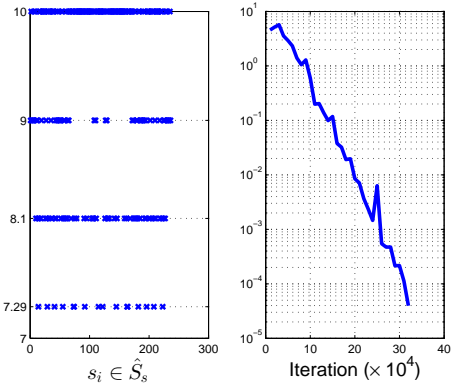}
\caption{Results for Example 1 for $N=4$: (Left) $\bar{J}^{\hat{\mathcal{G}}}_{\mathcal{R}}(\hat{\mu}, \hat{s})$ for all $\hat{s} \in \hat{S}_s$ and the learned greedy strategy $\hat{\mu}$; (right) the logarithm of the maximal change in $V$ in every $10^4$ iterations.}
\label{fig_demo1_result} \vspace{-0.2in}
\end{figure}

Now we illustrate the optimality of the learned greedy policy with the simulation result when $N=4$, whose result is shown in Fig.~\ref{fig_demo1_result}.
Let $\hat{\mu}$ be the greedy strategy of the system learned by the maximin-Q learning algorithm against an adversarial environment. 
If from a state $\hat{s} \in \hat{S}_s$ the system robot can only reach a diagonal position with respect to the position of the environment in at least $k \in \mathbb{N}$ steps, $\bar{J}^{\hat{\mathcal{G}}}_{\mathcal{R}}(\hat{\mu}', \hat{s})$ is upper bounded by $\sum_{l=k}^{\infty} \gamma^l \cdot 1 = \frac{1}{1-\gamma}\gamma^k$ for any system strategy $\hat{\mu}'$. By definition, if $\hat{\mu}^*$ is an optimal strategy for the system against an adversarial environment, we have $\bar{J}^{\hat{\mathcal{G}}}_{\mathcal{R}}(\hat{\mu}', \hat{s}) \leq \bar{J}^{\hat{\mathcal{G}}}_{\mathcal{R}}(\hat{\mu}^*, \hat{s}) \leq \frac{1}{1-\gamma}\gamma^k.$
In this 4-by-4 case, the system can always reach a diagonal position in 3 steps. Fig.~\ref{fig_demo1_result} shows that $V$ converges to the values 10, 9, 8.1 and 7.29, which coincide with $\frac{1}{1-\gamma}\gamma^k$ when $k=0,1,2,3$ and $\gamma=0.9$. Thus by the inequality above, $\bar{J}^{\hat{\mathcal{G}}}_{\mathcal{R}}(\hat{\mu}, \hat{s})$ also coincides with $\bar{J}^{\hat{\mathcal{G}}}_{\mathcal{R}}(\hat{\mu}^*, \hat{s})$, indicating that $\hat{\mu}$ itself is an optimal strategy of the system, as predicted by Algorithm~\ref{Algorithm_pseudo}.

{\bf Example 2:} Now we construct a new game $\mathcal{G}_1$ with a new winning condition $W_1 = (L, \varphi_1)$ from $\mathcal{G}$ by adding liveness assumptions to the environment robot and liveness requirements to the system robot. 
To be more specific, we require the system robot to visit the upper left corner (cell $N^2 - N$) and the lower right corner (cell $N-1$) infinitely often, provided that the environment robot visits the lower left corner (cell 0) and the upper right corner (cell $N^2-1$) infinitely often. $\mathcal{G}_1$ is the same as $\mathcal{G}$ except that $\varphi_1 = \varphi \wedge \big( (\square \diamondsuit x_{0} \wedge \square \diamondsuit x_{N^2 - 1}) \rightarrow (\square \diamondsuit y_{N-1} \wedge \square \diamondsuit y_{N^2 - N})\big)$.

\begin{figure}[ht!]
\vspace{-0.5\baselineskip}
\centering
\includegraphics[width=0.4\textwidth]{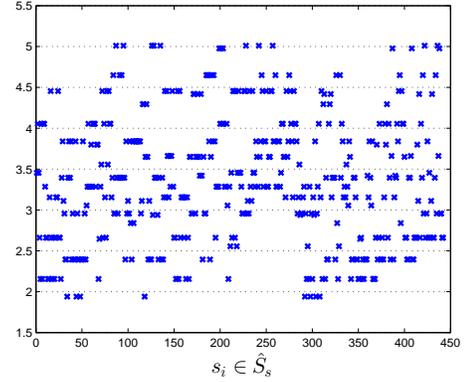}
\caption{Result for Example 2 when $N = 4$: $\bar{J}^{\hat{\mathcal{G}_1}}_{\mathcal{R}}(\hat{\mu}, \hat{s})$ for all $\hat{s} \in \hat{S}_s$ and a learned greedy strategy $\hat{\mu}$.}
\label{fig_demo2_result}
\end{figure}

The definition of the instantaneous function $\mathcal{R}$ remains the same as in Example 1, and the learning result of $\bar{J}^{\hat{\mathcal{G}_1}}_{\mathcal{R}}(\hat{\mu}, \hat{s})$ when $N = 4$ is given as Fig.~\ref{fig_demo2_result}. With this specification the system has no maximally permissive strategies, and it is expected that the true value of $\bar{J}^{\hat{\mathcal{G}_1}}_{\mathcal{R}}(\hat{\mu}, \hat{s})$ should be almost the same as $\bar{J}^{\hat{\mathcal{G}}}_{\mathcal{R}}(\hat{\mu}^*, \hat{s})$, as the system robot is allowed to follow $\hat{\mu}^*$ for as many finite moves as desired. However, Fig.~\ref{fig_demo2_result} shows that $\bar{J}^{\hat{\mathcal{G}_1}}_{\mathcal{R}}(\hat{\mu}, \hat{s})$ is smaller than $\bar{J}^{\hat{\mathcal{G}}}_{\mathcal{R}}(\hat{\mu}^*, \hat{s})$, indicating a sub-optimality due to the loss of some winning strategies by the permissive strategy.

{\bf Example 3:} We now illustrate the trade-off between the performance of the learned strategy and the computation cost in Algorithm~\ref{Algorithm_pseudo}. Consider a new game $\mathcal{G}_2$ with winning condition $W_2 = (L, \varphi_2)$ which is slightly different from the game $\mathcal{G}$ of Example 1 as it also requires the system robot to visit one of two given cells (say cell $N^2-N$ and cell $N-1$) infinitely often. In other words, $\varphi_2 = \varphi_0 \wedge \Box \diamondsuit (y_{N^2-N} \vee y_{N-1})$, which is in the form of GR(1). We compute a memoryless permissive strategy $\mu_2$ for $\mathcal{G}_2$. 

Now we design a sequence of games $\mathcal{G}_2^1$ to $\mathcal{G}_2^6$ from $\mathcal{G}$ in the following way. For each game we add a counter as a new controlled state variable which counts the number of the system's moves since its last visit to cell $N^2-N$ or cell $N-1$, but the maximum allowed counter values increases monotonically from $\mathcal{G}^1_2$ to $\mathcal{G}^6_2$. The value of each counter should always be less than its corresponding maximum value. All these 6 games satisfy the condition in Proposition~\ref{prop_special_safety} and we can extract a maximally permissive strategy for each of them. With the counters, the system robot is forced to visit cell $N^2-N$ or cell $N-1$ infinitely often and as a result, any permissive strategies of any game in this sequence is also a permissive strategy for $\mathcal{G}_2$. Let $\mu_2^i$ be the extracted maximally permissive strategy of the game $\mathcal{G}^i_2$ for $i = 1, \cdots, 6$. By definition of maximally permissive strategies, $\mu_2^i$ includes $\mu_2^j$ if $i > j$, $i,j \in \{1, \cdots, 6\}$. In this way we extracted a sequence of permissive strategies with increasing permissiveness for the game $\mathcal{G}_2$. 

We proceed the same learning procedure as the previous two examples on $\mathcal{G}_2$ and the game sequence from $\mathcal{G}^1_2$ to $\mathcal{G}^6_2$. For the 3-by-3 case, the maximum allowed counter values and the maximum values of the learned discounted reward are shown in Table~\ref{table_example3}. 
It is shown that the maximum discounted reward, which can be seen as the performance of the learned system strategy, increases monotonically with the maximum counter value, i.e., the permissiveness of the permissive strategy. In the meantime, the number of learning iterations and computation time grows. This illustrates the trade-off between the performance of the learned strategy and the computation cost. 

\begin{table}[h]
\centering
\caption{Results for example 3 (for the 3-by-3 case).}
\label{table_example3}
\begin{tabular}{| >{\centering\arraybackslash}m{1.1cm} | >{\centering\arraybackslash}m{1.2cm} | >{\centering\arraybackslash}m{1.6cm} | >{\centering\arraybackslash}m{1.2cm} | >{\centering\arraybackslash}m{1.4cm} |} 
\hline
{Strategy} & {Max counter value} & {Max discounted reward} & {Learning time [s]} & {Learning iterations [$\times 10^4$]}\\ \hline
$\mu_2$   &  N/A & 5.7368 & 9.87 & 20 \\ \hline 
$\mu_2^1$ &          4 & 8.0922 & 9.70 & 19 \\ \hline
$\mu_2^2$ &          6 & 8.8658 & 16.00 & 33 \\ \hline 
$\mu_2^3$ &          8 & 9.2442 & 27.34 & 55 \\ \hline 
$\mu_2^4$ &         10 & 9.4647 & 41.54 & 83 \\ \hline 
$\mu_2^5$ &         14 & 9.7034 & 83.88 & 172 \\ \hline
$\mu_2^6$ &         20 & 9.8616 & 275.88 & 534 \\ \hline
\end{tabular}
\end{table}
\vspace{-0.5\baselineskip}
\color{black}
\section{Concluding remarks}
We studied synthesis of optimal reactive controllers that are correct with respect to given temporal logic specifications. The performance criteria are unknown during design but can be inferred at run time. We proposed a solution that merges ideas from permissive strategy synthesis and reinforcement learning. We provided sufficient conditions (on the underlying temporal logic specifications) needed to acquire optimal performance, and demonstrated the algorithm on a number of robot motion planning examples.

\bibliographystyle{abbrv}
\bibliography{reference}

\end{document}